\documentstyle[prl,aps,epsf,floats]{revtex}

\begin{document}

\title{Quantum Suppression of the Rayleigh Instability in 
Nanowires}
\author{F.\ Kassubek\dag\ddag, 
C.\ A.\ Stafford\ddag, Hermann Grabert\dag\ and Raymond~E. Goldstein\ddag\S}
\address{\dag Fakult{\"a}t f{\"ur} Physik, Albert-Ludwigs-Universit{\"a}t,
Hermann-Herder-Stra{\ss}e 3, D-79104 Freiburg, Germany}
\address{\ddag Department of Physics and \S Program in Applied Mathematics,
University of Arizona, Tucson, AZ  85721}

\date{\today}

\maketitle

\begin{abstract}
A linear stability analysis of metallic nanowires is performed in the
free-electron model using quantum chaos techniques.  It is found that 
the classical instability of a long wire under surface tension
can be completely suppressed by electronic shell effects,
leading to stable cylindrical configurations whose electrical conductance 
is a magic number 1, 3, 5, 6,... times the quantum of conductance.
Our results are quantitatively consistent with recent
experiments with alkali metal nanowires.
\end{abstract}

\section{Introduction}

Plateau's 
celebrated study \cite{Plateau,chandrasekhar} of the stability of 
bodies under the influence of surface tension established a fundamental
result of classical continuum mechanics: a cylinder longer than its 
circumference
is energetically unstable to breakup.  Here we consider a quantum mechanical
generalization of Plateau's study---the stability of a metallic
nanowire.  Methods from the study of quantum chaos 
\cite{gutzwiller,SemiclPhys}
are used to examine the
energetics of a free electron gas in a cylindrical
jellium filament, and show that an 
infinite filament can be stable against all axisymmetric perturbations
if its electrical conductance $G$ [in units of 
$G_0=2e^2/h=(12,906.4 \, \mbox{Ohm})^{-1}$]
belongs to a set of ``magic numbers" 
$n=1,3,5,6,\ldots$ and is otherwise unstable. 
Here $e$ is the charge of an electron and $h$ is Planck's constant.
Such magic numbers are analogous to those
found in the shell model of atomic nuclei 
\cite{SemiclPhys} and in metal clusters \cite{SemiclPhys,clusters}.
Our stability analysis elucidates and confirms 
the stability properties of simple metal nanowires conjectured by
Yanson, Yanson, and van Ruitenbeek \cite{Yanson} 
based on their observation of shell structure in the
conductance statistics of sodium nanowires.

Structural multistability of metallic nanowires was
previously postulated based on classical molecular dynamics
simulations \cite{mol_dyn1,mol_dyn2,mol_dyn3}.
However, the magic numbers
observed in conductance histograms in alkali metals \cite{Yanson,Krans}
are clearly an electronic shell effect, as shown below, and 
cannot be explained with classical molecular dynamics.
The common occurence of multistability
in two such radically 
different models likely stems from the fact 
that both models
introduce an additional length scale (the Fermi wavelength $\lambda_F$ in the 
free electron
model, the atomic diameter in molecular dynamics simulations),
which leads to commensurability effects.  

The properties of nanowires formed from simple monovalent metals,
like the bulk properties of these materials \cite{Ashcroft+Mermin},
are determined largely by the delocalized conduction electrons.
A free electron jellium model,
treating the electrons as a noninteracting Fermi gas confined within the
wire by hard-wall boundary conditions,
provides an intuitive and even quantitative explanation of observed
quantities like conductance \cite{Torres,StaffordPRL97,BuerkiPRB99},
force 
\cite{StaffordPRL97,BuerkiPRB99,Jan97,Hoeppler99,StaffordPRL99}, 
and shot noise \cite{BuerkiPRL99}.
To examine energetic stability, 
we consider 
a weakly deformed cylinder, and find its thermodynamic potential 
to quadratic order in the amplitude of the deformation.  A contribution
proportional to the filament area, naturally identified as the surface
tension, is only one of several competing terms in the free energy, which
consists of terms which vary smoothly with the geometry of the filament
and an oscillatory contribution directly connected with the discrete energy
levels that are solutions to Schr\"odinger's equation.  
The smooth terms appear
in descending powers of length (proportional to volume, surface area, mean
curvature, etc.), and are quite analogous to those found 
in the study of classical wave equations in curved domains \cite{Balian} and
the related problem of classically screened Coulomb (Debye-H\"uckel) 
interactions
of curved surfaces in an electrolyte \cite{debyehuckel}.
The oscillatory part of the free energy, in particular, 
alters Plateau's classical stability analysis in an essential way.
The present work is thus fundamentally distinct from that which has dealt with 
the quantum mechanical origin of surface tension itself 
in metallic fluids \cite{ashcroft}, as well as those which consider
quantum mechanical corrections to classical surface tension
due to the quantization of capillary waves \cite{helium}.  

Long gold nanowires suspended between gold electrodes
have been produced and imaged with a transmission electron
microscope by Kondo and Takayanagi \cite{bridge,helix}; 
in particular, they have
observed wires which are stable and almost perfectly cylindrical.  
The wires of Ref.\ \cite{bridge,helix}
are of finite length, and attached to electrodes at either end.
It seems reasonable to consider an
infinite wire as a theoretical starting point, 
and to assume that the length of the wire acts 
as a cut-off: deformations with a wavelength longer than the wire length cannot
occur.  Thus one does not have to model the boundary conditions at the 
ends of the wire, which would introduce additional parameters.

\begin{figure}[t]
\epsfxsize=3.5truein
\centerline{\epsffile{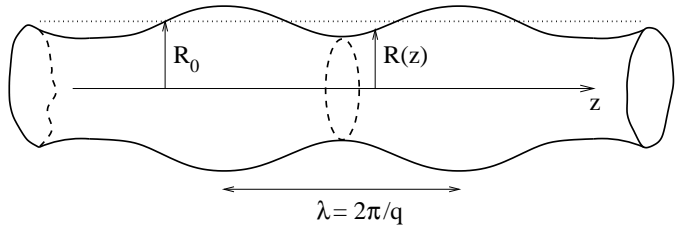}}
\smallskip
\caption{Deformation of wavevector $q$ of a cylindrical nanowire.}
\label{fig_deformation}
\end{figure}

To be specific, we examine an infinite cylindrical wire and its
sole classically unstable deformation---an 
axially symmetric one (see Fig.\ \ref{fig_deformation}). 
Any such deformation can be written as a Fourier series:
\begin{equation}
R(z)=R_0 \left(1+\int_0^\infty dq\, b(q) \cos( q z + \phi(q)) 
\right),
\label{eq:rofz}
\end{equation}
where $R(z)$ is the radius at position $z$, $R_0$ is the
unperturbed radius, 
$b(q)$ is the (infinitesimal) perturbation coefficient, and
$\phi(q)$ is an arbitrary phase shift.
The coefficients $b(q)$ are chosen such that
the total volume of the wire is unchanged by the deformation.
Other physically reasonable constraints \cite{StaffordPRL99}
are also possible, but lead to similar results.

The metallic nanowire is an open system, connected to 
macroscopic metallic electrodes at
each end \cite{Yanson,Krans,bridge,helix,Agrait,auhistogram,Auchain,Auchain2}. 
Therefore the change of the grand
canonical potential $\Omega$ under the perturbation determines its
stability. $\Omega$ 
is related to the electronic density of states $g(E)$ by
\begin{equation}
\label{OMEGA_ALS_FKT_V_G}
\Omega = - k_B T \int_0^\infty
 dE\, g(E) \ln \left(1+{\rm e}^{-(E-\mu)/k_BT} \right),
\end{equation}
where $k_B$ is Boltzmann's constant, $T$ is the temperature, and $\mu$
is the electronic chemical potential specified by the macroscopic electrodes.
Our aim is to expand $\Omega$ up to second order in the
coefficients $b(q)$ characterizing the deformation. As we will show,
this yields
\begin{equation}
\label{EXP_OMEGA}
\Omega[b]=\Omega[0]+\int_0^\infty\!\! dq\, \alpha(q) [b(q)]^2
+ {\cal{O}}(b^3)~.
\end{equation}
The change in the grand canonical potential is of second order in $b$
and contributions from deformations with different $q$ decouple. If
the prefactor $\alpha(q)$ is negative for any value of $q$, then $\Omega$
decreases under the deformation and the wire is unstable. 

\begin{figure}[t]
\epsfxsize=4truein
\centerline{\epsffile{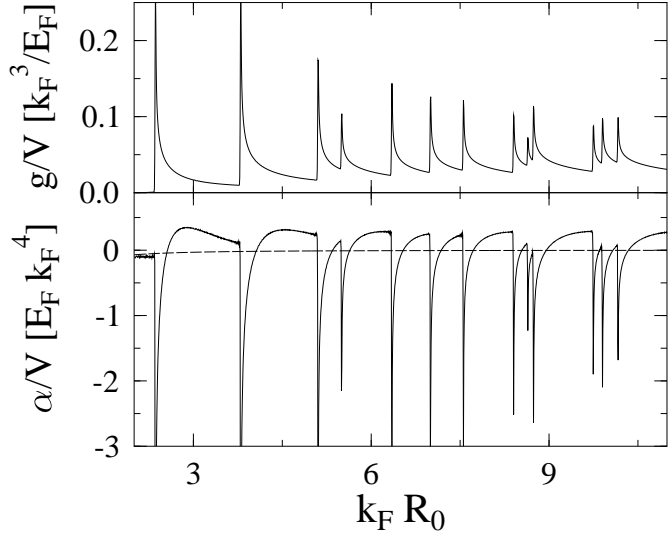}}
\smallskip
\caption{Density of states $g(E_F)$ of a cylindrical wire 
(upper diagram) and 
stability coefficient $\alpha$ (lower diagram) versus the 
radius $R_0$ of the unperturbed wire.
The wavevector of the perturbation is $q R_0 =1$.
Dashed curve: Weyl contribution to $\alpha$.
}
\label{fig1}
\end{figure}

For an infinite cylindrical wire, the transverse motion is quantized,
leading to the formation of discrete electronic subbands. The total density of
states is the sum of the contributions from each subband (see Fig.\ 
\ref{fig1}):  every subband
begins to contribute at a threshold energy equal to the energy of its 
quantized transverse motion with a sharp spike, falling 
off smoothly for increasing energy. 
If the Fermi energy $E_F$ lies near one of these sharp peaks,
certain small deformations of the wire can 
dramatically increase the density of
states. According to (\ref{OMEGA_ALS_FKT_V_G}), this lowers the grand
canonical potential, leading to an instability. On the other hand,
if there is no subband 
threshold sufficiently close to $E_F$, the density of states
will instead decrease with any deformation, implying the existence of
stable regions intervening between 
the instabilities associated with the opening of
each subband.

\section{Quantum Chaos Approach}

In order to examine this picture quantitatively, we use a
semiclassical approach \cite{gutzwiller,SemiclPhys}, 
which enables $g(E)$ to be split into a smooth average contribution
$\bar{g}(E)$, referred to as the {\em Weyl contribution},
and an oscillatory part $\delta g(E)$, whose average value 
is zero: 
\begin{equation}
g(E) = \bar{g}(E) + \delta g(E).
\end{equation}
The Weyl term
$\bar{g}$ contains terms proportional to the volume of the
nanowire, and to the area and curvature of its surface:
$\bar{g}(E)=(1/E_F)\tilde g(E/E_F)$, where
\begin{equation}
\label{DOS_WEYL}
\tilde g(x)= 
\frac{x^{1/2}}{2\pi^2} k_F^3 V 
-\frac{1}{8\pi} k_F^2 S
+ \frac{x^{-1/2}}{6\pi^2} k_F K,
\end{equation}
and $k_F=2\pi/\lambda_F$ is the Fermi wavevector.
The volume $V$, surface area $S$, and integrated mean curvature $K$ can be
calculated for arbitrary perturbations by simple geometric
considerations. 
The oscillatory part $\delta g$ 
of the density of states is a quantum correction,
and may be calculated in the semiclassical approximation as
a sum over all periodic classical orbits of the system
\cite{gutzwiller,SemiclPhys,Hoeppler99,StaffordPRL99,CreaghPRA90,UllmoPRE96,CreaghAnnPhys96}.

\begin{figure}[t]
\epsfxsize=3.5truein
\centerline{\epsffile{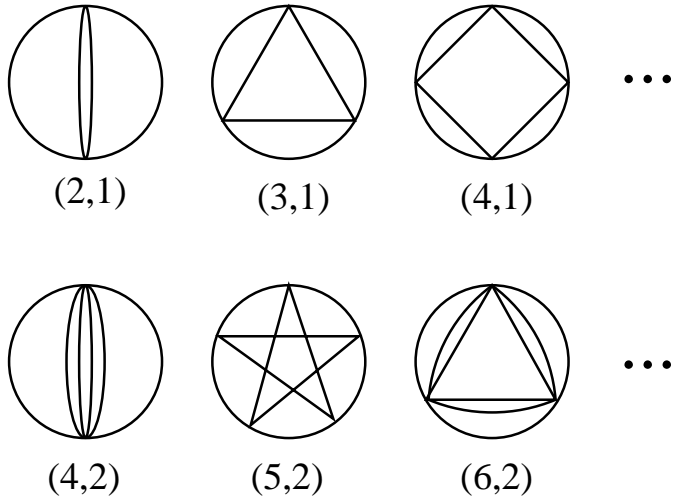}}
\bigskip
\bigskip
\caption{Classical periodic orbits of an electron in a plane perpendicular
to the $z$-axis, labeled $(v,w)$, where $v$ is the number of vertices
and $w$ the winding number.  
}
\label{fig_orbits}
\end{figure}

Let us first consider an undeformed cylindrical wire. Each periodic orbit 
lies in a plane transverse to the wire's axis,
and there is a correspondence to the orbits \cite{SemiclPhys} 
in a circular billiard (see Fig.\ \ref{fig_orbits}).  Note that
the action $S_{vw}$ of each orbit is invariant under both translations of the
orbit parallel to the $z$-axis and rotations about the $z$-axis.
In a system with continuous symmetries, when taking the trace of the 
electron Green's function to calculate the density of states, one must
first integrate exactly over these symmetries \cite{SemiclPhys,CreaghPRA90}
before employing the semiclassical (stationary phase) approximation.
For electrons in a cylinder of length $L$, 
one obtains the following trace formula 
\begin{equation}
\delta g(E)  = 
\frac{mL}{\pi \hbar^2} 
\sum_{w=1}^\infty \sum_{v=2w}^\infty \frac{f_{vw} L_{vw}}{v^2}
\cos\left(\frac{S_{vw}(E)}{\hbar}- \frac{3v\pi}{2} \right),
\label{DOS_GUTZ}
\end{equation}
where $m$ is the electron mass, $v$ and $w$ are defined in Fig.\ 
\ref{fig_orbits}, 
$f_{vw}=1+\theta(v-2w)$ counts the discrete degeneracy of the orbit under
time reversal,
$L_{vw}= 2vR_0 \sin(\pi w/v)$ is the length of a periodic orbit, and
the action $S_{vw}/\hbar=k_F L_{vw}\sqrt{E/E_F}$.

When the wire is
deformed, the translational symmetry is broken, and we use
a semiclassical perturbation theory \cite{UllmoPRE96,CreaghAnnPhys96}
to calculate $\delta g$.
This theory assumes that for small
perturbations, the  amplitudes and topology of the unperturbed orbits 
can still
be used in (\ref{DOS_GUTZ}), but that the lengths of the orbits---and hence 
their actions---change. 
The trace formula becomes
\begin{equation}
\delta g(E)  = 
\frac{m}{\pi \hbar^2} 
\sum_{w=1}^\infty \sum_{v=2w}^\infty \frac{f_{vw}}{v^2}
\mbox{Re} \left\{
e^{i(S_{vw}/\hbar- 3v\pi/2)} 
\int_0^L dz \,(L_{vw}+\Delta L_{vw}) e^{i\Delta S_{vw}/\hbar}
\right\},
\label{DOS_PERT}
\end{equation}
where $\Delta S_{vw}/\hbar=k_F \Delta L_{vw}\sqrt{E/E_F}$, and 
\begin{equation}
\Delta L_{vw} = 2 v \sin(\pi w/v)(R(z) - R_0),
\end{equation}
which may be expressed in terms of the perturbation coefficients using Eq.\
(\ref{eq:rofz}).
Combining this result for $\delta g$ with
(\ref{DOS_WEYL}), it is straightforward to
calculate the density of states up to second order in the coefficients
$b(q)$ for a deformed wire. The result can be integrated to obtain the
grand canonical potential, which indeed has the expansion (\ref{EXP_OMEGA}).

\section{Stability Analysis}

Let us first discuss the stability of a nanowire at zero temperature. 
Fig.\ \ref{fig1} shows the stability coefficient $\alpha$ (lower diagram)
and the density of states at the Fermi energy (upper diagram) as functions
of the radius of the unperturbed wire.  The wavelength of the perturbation
was taken to be $q R_0=1$, the critical wavelength for stability 
in Plateau's classical analysis of a body under the
influence of surface tension.  In addition to surface tension, the present
model for $\Omega$
has a curvature energy, which enhances the instability for small
$R_0$, and an oscillatory component associated with the opening of 
successive subbands as $R_0$ increases.  As discussed above, $\alpha$
has sharp negative peaks---indicating strong instabilities---at 
the subband thresholds, where the density of states is sharply peaked.
Under surface tension and curvature energy alone
(dashed curve in Fig.\ \ref{fig1}), the wire 
would be slightly unstable at the critical
wavevector $q R_0=1$, since the curvature term is negative.
However, the quantum correction is positive in the
regions between the thresholds to open new subbands, {\em
thus stabilizing the wire}.  Since the oscillatory contribution to $\alpha$
is independent of $q$, we find that regions of stability persist 
for {\em arbitrarilly long wavelength perturbations}, indicating that 
an infinitely long cylindrical wire is a true metastable state if the radius
lies in one of the windows of stability.

\begin{figure}[t]
\epsfxsize=4.0truein
\centerline{\epsffile{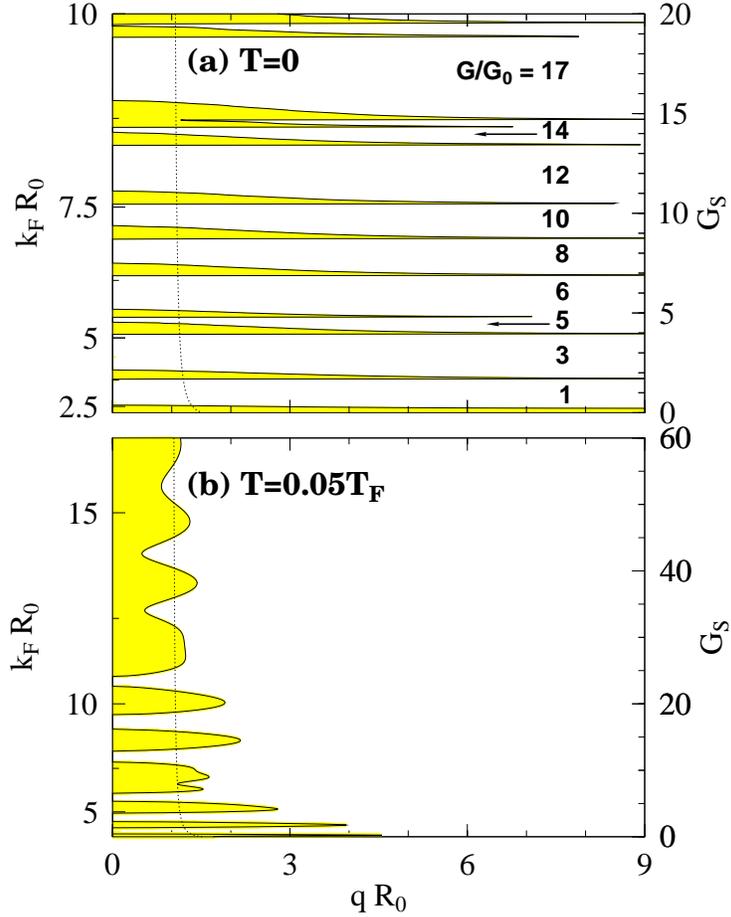}}
\bigskip
\caption{Stability diagram for cylindrical nanowires at two different 
temperatures.
White areas are stable, grey unstable to small perturbations.
The quantized electrical conductance values $G$ of the stable 
configurations are indicated by bold numerals in (a), with $G_0=2e^2/h$. 
Right vertical axis: corrected Sharvin conductance $G_S$.
Dotted curve: stability criterion in the Weyl approximation.
}
\label{fig2}
\end{figure}

With these results, we can construct the zero temperature
stability diagram for the wire
[see Fig.\ \ref{fig2}(a)].  
In contrast to Plateau's classical stability
analysis, an additional quantum length scale 
arises here, namely the Fermi wavelength $\lambda_F$.  The stability problem
is now determined by two dimensionless parameters: $qR_0$ and 
$k_F R_0$.  In Fig.\ \ref{fig2},
regions of instability, 
where the coefficient $\alpha(q)$ is negative, are
shaded grey,  while the stable regions are shown in white.
Note that many of the white regions of stability persist 
all the way down to $q=0$.  
The {\em multistability} of the system,
indicated by the alternating stable and unstable stripes, reflects
commensurability effects between $\lambda_F$ and $R_0$.

We remark that in addition to the axially symmetric modes considered here,
which are the sole unstable modes in the classical limit, perturbations
which break axial symmetry may also become unstable near the subband 
thresholds.  However, these Jahn-Teller like modes \cite{Jan97}
are likely less unstable
than the axisymmetric modes, and will not destroy the regions of stability
shown in Fig.\ \ref{fig2}(a).

\section{Conductance Magic Numbers}

The electrical conductance $G$ of a perfect cylindrical nanowire 
is quantized \cite{Torres} in units of $G_0=2e^2/h$.
The quantized conductance values of the stable cylindrical
configurations are indicated by bold numerals in Fig.\ \ref{fig2}(a). 
For comparison, the right vertical axis of the figure 
is labeled with the corrected Sharvin conductance \cite{Torres} 
\begin{equation}
G_S =\left(\frac{k_F R_0}{2}\right)^2
\left(1 -\frac{2}{k_F R_0}\right),
\label{eq:Sharvin}
\end{equation}
which gives a smooth approximation to $G/G_0$. 
The conductance values of the stable configurations
are somewhat analogous to the {\em magic numbers} of enhanced stability
in atomic nuclei \cite{SemiclPhys} and metal clusters 
\cite{SemiclPhys,clusters}.  An important distinction is that the magic numbers
in nuclei and clusters refer to the number of fermions in a finite system,
while we consider an infinite, open system, with magic numbers describing
the number of conducting transverse modes \cite{StaffordPRL97} 
which hold the wire together.   The number of conducting modes is approximately
equal to the number of atoms which fit within a cross section of the wire.

The integer magic numbers identified here must be distinguished 
from the concept of ``magic wire configurations'' proposed by van Ruitenbeek
and others \cite{Jan97,yannouleas,zabala}.
The latter represented discrete minima of
the energy per unit volume of a jellium wire 
as a function of its cross-sectional area,
but the stability with respect to deformations with $q > 0$
was not considered.
A similar scenario was advanced in Ref.\ 
\cite{Yanson} based on a partial summation of the periodic orbits in the
expression for $\Omega$ itself.  While
the theoretical results of Refs.\ \cite{Yanson,Jan97,yannouleas,zabala} 
seem to imply
that stability occurs 
only for a discrete set of radii, of measure zero, our
analysis finds stability with respect to small perturbations
over broad intervals of radius. 
Furthermore, the very existence of discrete energetic minima of the type 
discussed in Refs.\ \cite{Yanson,Jan97,yannouleas,zabala} depends sensitively
on the numerical value of the surface tension in the model; for example, 
they do not occur whatsoever in the free-electron model at constant volume
for $G>G_0$.  In contrast, the finite windows of stability in our analysis
are robust with respect to variations of the surface tension.

\section{Comparison to Experiments in Alkali Metals}

The sequence of magic numbers $G/G_0=1, 3, 5, 6, \ldots$ is consistent
with the observation of conductance quantization
in alkali metal point contacts \cite{Krans}.
Recently, conductance histograms for sodium nanowires
with pronounced peaks up to $G/G_0 \sim 100$ were obtained by Yanson 
{\em et al.} \cite{Yanson}.  They argued that these peaks could
not be understood based on conductance quantization alone, but rather 
reflected energetically preferred wire configurations.
In order to construct a theoretical conductance histogram, we need to
know the {\em a priori probability} of occurence of a nanowire of a given
cross-sectional area.  Here, we make the simplest 
hypothesis: that {\em nanowires of different cross-sectional
areas occur with equal a priori probability 
if stable, and with zero probability otherwise}.
On this hypothesis, the relative probability of observing a contact with
a given quantized conductance value is proportional to
the width $\Delta G_S$ of the corresponding stable region,
$G_S$ being a dimensionless measure of the contact area.

\begin{figure}[t]
\epsfxsize=4.0truein
\centerline{\epsffile{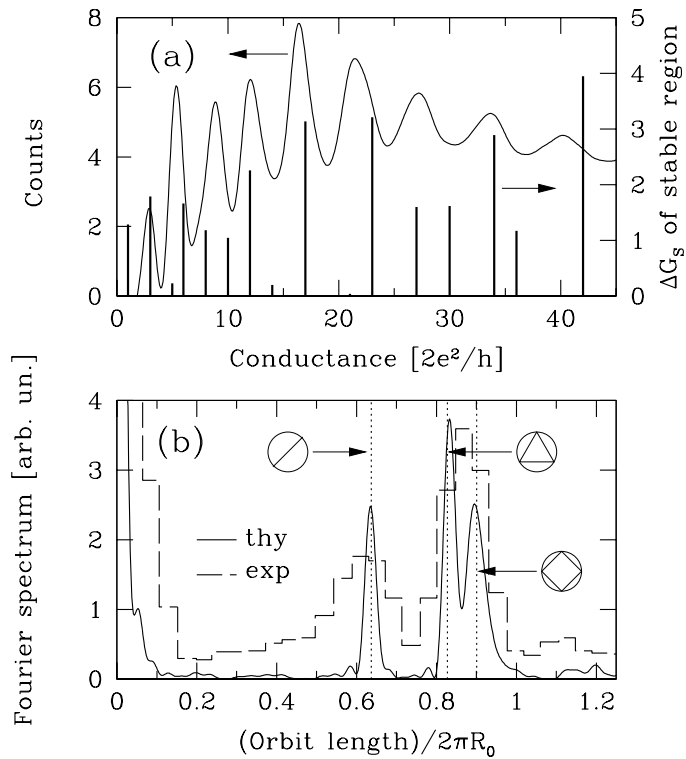}}
\bigskip
\caption{
Magic numbers: theory and experiment. (a)
Solid curve: conductance histogram for sodium nanowires at $T=80\mbox{K}
=0.002T_F$,
reproduced from Ref.\ \protect\cite{Yanson}.  
Vertical bars: quantized conductance
values of the metastable nanowire configurations in the free electron model
at the same temperature.  The height of each bar is equal to the width 
$\Delta G_S$ of the corresponding stable region 
[c.f.\ Fig.\ \ref{fig2}(a)]. (b) Solid curve: 
Fourier power spectrum of the theoretical
conductance histogram [vertical bars in (a)], displaying the dominant
contributions of the three shortest periodic orbits in Eq.\
(\ref{DOS_PERT}).  Dashed curve: Fourier power spectrum of the conductance
histogram for sodium nanowires at $T=90\mbox{K}$, reproduced from
Ref.\ \protect\cite{Yanson00}.  Note that the theoretical Fourier spectrum
does change significantly between 80K and 90K.
}
\label{fig3}
\end{figure}

The conductance histogram from Ref.\ \cite{Yanson} taken at a temperature
of $T=80\mbox{K}$ is reproduced in our Fig.\ \ref{fig3}(a).  
For comparison, the theoretical magic numbers at the
same temperature are plotted as vertical bars, with height equal to
the width $\Delta G_S$ of the corresponding stable region.
Unlike the idealized wires in our analysis,
the experimental wires are of finite length, and contain imperfections.
Thus the peaks
in the experimental histogram are shifted \cite{BuerkiPRB99}   
below the theoretical integer values due to backscattering, 
and are broadened \cite{BuerkiPRB99} due to tunneling, 
disorder-induced conductance fluctuations, and inelastic processes.
Furthermore, the relative heights
of the peaks may be influenced by dynamical as well as energetic effects;  
in particular, the peak at $G=G_0$, which is quite pronounced at the 
lowest temperatures \cite{Yanson,Krans}, is absent from the experimental
histogram at $T=80\mbox{K}$---presumably indicating that thermally activated
processes lead to the rupture of this metastable configuration.

While the first four theoretical peaks at $G/G_0=1,3,5,6$
can be identified unambiguously with the narrower
peaks in the low temperature data of Refs.\ \cite{Yanson,Krans},
it is not entirely clear whether one can match up the broad peaks
in the 80K experimental histogram with individual theoretical magic 
numbers.  In particular, there 
appears to be
fine structure in the 
theoretical histogram which is not present in the 80K histogram,
possibly because it is obscured due to broadening.
In order to see whether the theoretical histogram nonetheless correctly
describes the gross features of the experimental histogram, it is 
useful to take the Fourier transforms of the two histograms, singling out the
contributions of the shortest periodic orbits. 

The actions of
the periodic orbits are proportional to the radius of the wire, and hence
are approximately proportional to the square root of the conductance,
by Eq.\ (\ref{eq:Sharvin}).  The contributions of the various periodic
orbits to the conductance histogram can thus be extracted \cite{Yanson00}
by taking a Fourier transform with respect to $\sqrt{G/G_0}$.
The Fourier power spectrum of the theoretical conductance histogram is
shown as a solid curve 
in Fig.\ \ref{fig3}(b), where all magic numbers up to $G/G_0=207$ were 
included.  It shows clear peaks corresponding to the
three shortest periodic orbits.  For comparison, 
the experimental Fourier spectrum for sodium nanowires at 90K obtained by
Yanson {\it et al.} \cite{Yanson00} is shown as a dashed curve.  
The experimental spectrum contains two broad peaks, one centered at the
frequency of the diameter orbit, and a second which spans the frequencies
of the triangle and square orbits.
The experimental peaks in Fig.\ \ref{fig3}(b) 
are broader than the theoretical peaks since the oscillatory 
structure in the experimental histogram is damped for $G/G_0 > 50$.
The widths of the theoretical peaks, in contrast, are 
determined by the maximum conductance included, 
$\mbox{max}\{G/G_0\}=207$; increasing this cutoff decreases the widths of
the peaks.
We have simply taken a cutoff larger than the 
maximum measured conductance, so that the cutoff does not introduce any 
artifacts in the theoretical Fourier spectrum.
The overall vertical scale of the power spectra shown is arbitrary, so 
a comparison between theory and experiment must be based on the 
relative spectral weights of the various peaks.
In both spectra, the weight of the diameter orbit is roughly half
the combined spectral weights of the triangle and square orbits---a
rather good agreement between theory and experiment.

\section{Classical Limit}

At zero temperature, the pattern of stable regions separated by unstable
stripes shown in Fig.\ \ref{fig2}(a) continues up to arbitrarilly large radii.
However, at any finite temperature $T$, the quantum oscillations in $\alpha$
are smoothed out, and the classical stability criterion $qR_0>1$ is recovered
asymptotically
for sufficiently large radii.  The crossover from the $T=0$ result to the
classical limit occurs when $k_B T \sim E_F (G_0/G)$,
i.e., when the thermal
energy $k_B T$ is comparable to the average transverse level spacing.
Fig.\ \ref{fig2}(b) shows the stability diagram for $T/T_F=.05$,
where $T_F=E_F/k_B$ is the Fermi temperature.
One sees that the stability boundary indeed 
begins to cross over to the classical
line $q R_0=1$ for $G_S > 20$.  

In Fig.\ \ref{fig2}(b), 
there are no true metastable configurations
with $G_S>25\sim T_F/T$,
indicating that all thicker wires 
would be dynamically unstable (like a column of fluid)  
at this temperature, once
the electronic shell effects have been smoothed out.
However, $T_F = 3.75\times 10^4 \mbox{K}$ in sodium, so multistability from 
electronic
shell effects can be expected to occur in sodium contacts with $G/G_0 
\lesssim 125$ up to at least 300K.

\section{Discussion}

It should be pointed out that thermal averaging is not the 
only mechanism which can suppress electronic shell effects.  Disorder 
also tends to smooth out the sharply peaked structure in the density of states,
so that one can expect a reduction of shell effects when the diameter of the 
wire exceeds the mean free path.  Furthermore, the tendency of the positive
ions to order themselves into regular arrays \cite{bridge,helix}
will certainly affect the stability of metallic nanowires.  Indeed, pioneering
theoretical investigations \cite{mol_dyn1,mol_dyn2,mol_dyn3} 
of the dynamics of nanowires focused exclusively on 
the arrangement of the ions.  Based on the 
relative importance of electronic shells and crystal structure
in metal clusters \cite{clusters}, 
one would expect electronic shell effects to dominate 
the energetics of very thin wires, particularly in the alkali metals,
with crystal structure becoming increasingly important for thicker
wires, and for metals where the bonding is more directional.

Although conductance histograms of
gold nanowires \cite{auhistogram}
do not exhibit the sequence of magic numbers 
predicted here and observed in alkali metals, there is some evidence
that gold nanowires can otherwise be adequately described using the free 
electron model \cite{StaffordPRL97,BuerkiPRB99,BuerkiPRL99}.
It is thus worthwhile to speculate about a possible electronic origin for 
the remarkable stability of wires of individual gold atoms. 
Linear chains composed of four to seven 
gold atoms suspended between two gold electrodes, with a conductance $G=G_0$,
were found to be stable in the laboratory for hours at a time 
\cite{Auchain,Auchain2}.
Given that such a configuration
has an enormous surface energy, its stability is at first sight surprising.
However, in our free electron model, we find that an infinitely long
wire with a conductance of $G_0$ is indeed stable with respect to small
perturbations.

Our results may also be relevant to explain the observed stability of 
gold wires with larger radii \cite{bridge,helix}.
The atomic-scale structure of these wires exhibits a remarkable diversity,
ranging from helical, multishell structures \cite{helix}
to crystalline structures with surface reconstruction \cite{bridge}; but 
a common factor in the observed stable structures is that they are 
almost perfectly cylindrical.  
A direct comparison to our theoretical
stability analysis would be facilitated by measurements 
of the conductance of the stable wires, which have not yet been carried out.
However, a rule of thumb is that the conductance $G/G_0$
of a metal wire made of monovalent atoms
is roughly equal to the number of atoms in the cross section. 
Thus the hexagonal prism structure with a cross
section of 30 atoms determined in Ref.\ \cite{bridge} does correspond to
a stable configuration in our analysis [vertical bar at $G/G_0=30$ in
Fig.\ \ref{fig3}(a)].  Similarly, the thinnest of the helical wire structures 
determined in Ref.\ \cite{helix}, with a cross section of 8 atoms, also 
corresponds to a stable structure in our analysis, while the two thickest
wire structures determined in Ref.\ \cite{helix}, with cross sections of
22 and 24 atoms, respectively, appear to straddle the predicted island of
stability at $G/G_0=23$.  Further work is needed to elucidate the 
interplay of atomistic stacking effects and electronic shell effects in these
structures.

Finally, let us comment on the dynamical evolution
of a nanowire under elongation or compression.  Consider stretching a
nanowire that is initially in a stable configuration (white areas in Fig.\
\ref{fig2}).
Under elongation, the radius of the wire decreases, so that one moves
downward on the stability diagram.  When a stability
boundary is encountered, it becomes energetically (and dynamically)
favorable for the wire to deform spontaneously, until 
another stable configuration of smaller radius is reached, thus
causing the conductance to 
jump abruptly from one magic number to a smaller one, and conversely under
compression.  This scenario is consistent with the claim \cite{Agrait}
that the structure of a metallic nanowire 
undergoes a sequence of abrupt changes as a function of
elongation or compression. 
The finite widths of the unstable tongues in Fig.\ \ref{fig2}
also provides a possible explanation for the hysteresis
\cite{Agrait} observed in the conductance as a function of
elongation: the critical radius at which the wire's conductance jumps between
neighboring magic numbers is different, 
depending on whether the tongue is approached from above or below, i.e., 
depending on whether the wire is stretched or compressed. 

\section*{Acknowledgments}

After completion of this work, we became aware of additional relevant
experimental results and a novel data analysis by A.\ I.\ Yanson
{\em et al.} \cite{Yanson00}.
We thank J.\ M.\ van Ruitenbeek and A.\ I.\ Yanson
for extensive discussions, and for permission
to reprint the experimental data shown in Fig.\ \ref{fig3}.
F.K.\ and H.G.\ were supported by Grant
SFB 276 of the Deutsche Forschungsgemeinschaft, 
C.A.S.\ by NSF Grant DMR0072703,
and R.E.G.\ by NSF Grant DMR9812526.
This research was supported by an award from Research Corporation.


\begin{thebibliography}{00}

\bibitem{Plateau}
Plateau J 1873 {\em Statique exp\'erimentale
et th\'eorique des liquides soumis aux seules forces mol\'eculaires},
(Paris: Gautier-Villars).

\bibitem{chandrasekhar}
Chandrasekhar S 1981 {\em Hydrodynamic and Hydromagnetic Stability}
(New York: Dover), pp. 515-574.

\bibitem{gutzwiller} Gutzwiller M C 1990 {\em Chaos in Classical and Quantum
Mechanics} (New York: Springer-Verlag).

\bibitem{SemiclPhys}
Brack M and Bhaduri R K 1997
{\em Semiclassical Physics}
(Reading, MA: Addison-Wesley).

\bibitem{clusters} 
de Heer W A 1993 The physics of simple metal clusters: experimental aspects 
and simple models {\em Rev. Mod. Phys.} {\bf 65}, 611-676. 

\bibitem{Yanson}
Yanson A I, Yanson I K and van Ruitenbeek J M 1999
Observation of shell structure in sodium nanowires
{\em Nature} {\bf 400}, 144-146.

\bibitem{mol_dyn1}
Landman U, Luedtke W D, Burnham N A and Colton R J 1990 
Atomistic mechanisms and dynamics of adhesion, nanoindentation, and fracture
{\em Science} {\bf 248}, 454-461.

\bibitem{mol_dyn2}
Todorov T N and Sutton A P 1993
Jumps in electronic conductance due to mechanical instabilities
{\em Phys. Rev. Lett.} {\bf 70}, 2138-2141.

\bibitem{mol_dyn3} 
S\o rensen M R, Brandbyge M and Jacobsen K W 1998
Mechanical deformation of atomic-scale metallic contacts: Structure and
mechanisms
{\em Phys. Rev. B} {\bf 57}, 3283-3294.

\bibitem{Krans}
Krans J M, van Ruitenbeek J M, Fisun V V, Yanson I K and
de Jongh L J 1995 
The signature of conductance quantization in
metallic point contacts
{\em Nature} {\bf 375}, 767-769.

\bibitem{Ashcroft+Mermin}
Ashcroft N W and Mermin N D 1976 {\em Solid State Physics}
(New York: Saunders College Publishing), pp. 29-55.

\bibitem{Torres}
Torres J A, Pascual J I and S\'aenz J J 1994
Theory of conduction through narrow constrictions in a three-dimensional
electron gas
{\em Phys. Rev. B} {\bf 49}, 16581-16584.

\bibitem{StaffordPRL97}
Stafford C A, Baeriswyl D and B{\"u}rki J 1997
Jellium model of metallic nanocohesion
{\em Phys. Rev. Lett.} {\bf 79}, 2863-2866. 

\bibitem{BuerkiPRB99}
B\"urki J, Stafford C A, Zotos X and Baeriswyl D 1999 
Cohesion and conductance of disordered metallic point contacts
{\em Phys. Rev. B} {\bf 60}, 5000-5008.

\bibitem{Jan97}
van Ruitenbeek J M, Devoret M H, Esteve D and Urbina C 1997 
Conductance quantization in metals: The influence of subband formation on the 
relative stability of specific contact diameters
{\em Phys. Rev. B} {\bf 56}, 12566-12572.

\bibitem{Hoeppler99}
H\"oppler C and Zwerger W 1999 
Quantum fluctuations in the cohesive force of metallic nanowires
{\em Phys. Rev. B} {\bf 59}, R7849-R7851. 

\bibitem{StaffordPRL99}
Stafford C A, Kassubek F, B{\"u}rki J and Grabert H 1999
Universality in Metallic Nanocohesion: A Quantum Chaos Approach
{\em Phys. Rev. Lett.} {\bf 83}, 4836-4839. 

\bibitem{BuerkiPRL99} 
B\"urki J and Stafford C A 1999 
Comment on ``Quantum Suppression of Shot Noise in Atom-Size Metallic Contacts''
{\em Phys. Rev. Lett.} {\bf 83}, 3342.

\bibitem{Balian}
Balian R and Bloch C 1970 
Distribution of eigenfrequencies for the wave equation in a finite domain I:
3D problem with a smooth boundary surface
{\em Ann. Phys. (N.Y.)} {\bf 60}, 401-447.

\bibitem{debyehuckel}
Duplantier B, Goldstein R E, Romero-Roch\'{\i}n V and Pesci A I 1990
Geometrical and topological aspects of electric double layers
near curved surfaces
{\em Phys. Rev. Lett.} {\bf 65}, 508-511.

\bibitem{ashcroft}
D'Evelyn M P and Rice S A 1983
Pseudoatom theory for the liquid-vapor
interface of a simple metal: computer simulation studies of sodium and
cesium
{\em J. Chem. Phys.} {\bf 78}, 5225-5249.

\bibitem{helium}
Ebner C and Saam W F 1975 
Renormalized density-functional theory of nonuniform superfluid
$^4$He at zero temperature
{\em Phys. Rev. B} {\bf 12}, 923-939.

\bibitem{bridge}
Kondo Y and Takayanagi K 1997
Gold Nanobridge Stabilized by Surface Structure
{\em Phys. Rev. Lett.} {\bf 79}, 3455-3458.

\bibitem{helix}
Kondo Y and Takayanagi K 2000
Synthesis and Characterization of Helical Multi-Shell Gold Nanowires
{\em Science} {\bf 289}, 606-608.

\bibitem{Agrait} 
Rubio G, Agra\"{\i}t N and Vieira S 1996  
Atomic-Sized Metallic Contacts: Mechanical Properties and Electronic Transport
{\em Phys. Rev. Lett.} {\bf 76}, 2302-2305.

\bibitem{auhistogram}
Costa-Kr\"amer J L, Garc\'{\i}a N and Olin H 1997
Conductance quantization histograms of gold nanowires at 4 K
{\em Phys. Rev. B} {\bf 55}, 12910-12913.

\bibitem{Auchain} 
Ohnishi H, Kondo Y and Takayanagi K 1999
Quantized conductance through individual rows of suspended gold atoms
{\em Nature} {\bf 395}, 780-783.

\bibitem{Auchain2}
Yanson A I, Rubio Bollinger G, van den Brom H E, Agra\"{\i}t N
and van Ruitenbeek J M
Formation and manipulation of a metallic wire of single gold atoms
(1999) {\em Nature} {\bf 395}, 783-785.

\bibitem{CreaghPRA90}
Creagh S C and Littlejohn R G 1991 
Semiclassical trace formulas in the presence of continuous symmetries
{\em Phys. Rev. A} {\bf 44}, 836-850.

\bibitem{UllmoPRE96}
Ullmo D, Grinberg M and Tomsovic S 1996   
Near-integrable systems: Resonances and semiclassical trace formulas
{\em Phys. Rev. E} {\bf 54}, 136-152.

\bibitem{CreaghAnnPhys96}
Creagh S C 1996
Trace formula for broken symmetry
{\em Ann. Phys. (N.Y.)} {\bf 248}, 60-94.

\bibitem{yannouleas} 
Yannouleas C and Landman U 1997
On Mesoscopic Forces and Quantized Conductance in Model Metallic Nanowires
{\em J. Phys. Chem. B} {\bf 101}, 5780-5783.

\bibitem{zabala}
Zabala N, Puska M J and Nieminen R M 1999 
Electronic structure of cylindrical simple-metal nanowires in the
stabilized jellium model
{\em Phys. Rev. B} {\bf 59}, 12652-12660.

\bibitem{Yanson00}
Yanson A I, Yanson I K and van Ruitenbeek J M 2000 
Supershell Structure in Alkali Metal Nanowires
{\em Phys. Rev. Lett.} {\bf 84}, 5832-5835.

\end{thebibliography}
\end{document}